\documentclass[%
 reprint,
superscriptaddress,
showpacs,preprintnumbers,
amsmath,amssymb,
aps,
prb,
]{revtex4-2}

\usepackage{graphicx}% Include figure files
\usepackage{dcolumn}% Align table columns on decimal point
\usepackage{bm}% bold math
\usepackage{amsmath}
\begin{document}

%\preprint{APS/123-QED}

\title{Tunable edge and depth sensing via phase-change nonlocal metasurfaces}
% Force line breaks with \\

\author{Kenan Guo}
\affiliation{School of Information Engineering, Nanchang University, Nanchang 330031, China}
\affiliation{Institute for Advanced Study, Nanchang University, Nanchang 330031, China}

\author{Yue Jiang}
\affiliation{Jiluan Academy, Nanchang University, Nanchang 330031, China}

\author{Shuyuan Xiao}
\email{syxiao@ncu.edu.cn}
\affiliation{School of Information Engineering, Nanchang University, Nanchang 330031, China}
\affiliation{Institute for Advanced Study, Nanchang University, Nanchang 330031, China}

\author{Tingting Liu}
\email{ttliu@ncu.edu.cn}
\affiliation{School of Information Engineering, Nanchang University, Nanchang 330031, China}
\affiliation{Institute for Advanced Study, Nanchang University, Nanchang 330031, China}

\begin{abstract}
Performing simultaneous depth-of-field (DoF) extension and edge enhancement within a single optical element remains a fundamental challenge in advanced imaging. Here, we propose a wavelength-tunable nonlocal Huygens' metasurface capable of simultaneously extracting depth and edge features of images in a single-shot exposure. Using the selective polarization response of the Huygens' metasurfaces, the circularly polarized converted component undergoes geometric phase modulation for wavefront shaping to extend the DoF, while the non-converted component acts as a spatial frequency filter to enhance edge contrast. The integration of a phase-change material, Sb$_{2}$S$_{3}$, enables continuous tuning of the resonance wavelength across a range of 100 nm by modulating its refractive index, granting the system excellent broadband spectral adaptability. This work offers a novel and compact solution for real-time depth sensing and feature extraction in applications such as autonomous navigation and biomedical imaging.
% \begin{description}
% Secondary publications and information retrieval purposes.
% You may use the \texttt{description} environment to structure your abstract;
% use the optional argument of the \verb+\item+ command to give the category of each item. 
% \end{description}
\end{abstract}

\maketitle

%\tableofcontents

\section{Introduction}

The integration of edge detection and depth information extraction provides critical morphological features and stereoscopic distance metrics, enhancing the visualization and analysis of three-dimensional (3D) objects for applications such as high-precision facial biometrics {\cite{kim2021nanophotonics,chen2023meta,hsu2024metasurface}}, surgical navigation and perception of autonomous vehicles {\cite{fujiyoshi2019deep,chen2022meta}}. Traditional depth extraction requires real-time electromechanical tuning for multi-frame defocused acquisition, while classical edge detection typically employs a 4f optical system for momentum-space filtering {\cite{goodman2005introduction,abdollahramezani2020meta}}. These discrete architectures suffer from high complexity and large physical footprints, making them ill-suited for the miniaturization and integration of modern optical systems. Optical metasurfaces offer a promising alternative through precise wavefront control over phase, amplitude, and polarization within ultrathin platforms {\cite{tittl2018imaging,huang2018metasurface,ding2020metasurface,zhao2020recent,tseng2020dielectric}}. In previous work, metasurface designs using a double-helix point spread function (DH-PSF) enable sub-millimeter depth accuracy via rotating foci {\cite{Grover2010,conkey2011,quirin2012,shen2023monocular}}, while those employing the photonic spin Hall effect achieve edge extraction {\cite{zhou2019,li2019}}. Advancing beyond single-function devices, multi-dimensional multiplexing strategies facilitate simultaneous depth mapping and edge detection via vortex beam modulation within single nanostructures {\cite{yang2023}}. Furthermore, a similar multiplexing metasurface mechanism has been extended to achieve multifunctionality such as differential and bright-field imaging {\cite{yang2021}}. Despite these advances, most of these local metasurface designs rely on unit-cell spatial modulation to shape wavefronts over broad spectra, which inherently lack wavelength selectivity and inevitably introduce wavefront disturbances at non-target wavelengths, fundamentally hindering performance optimization in multi-wavelength multiplexed imaging scenarios.

Compared with conventional local metasurfaces, nonlocal metasurfaces enable simultaneous spectral and spatial modulation, offering inherent high frequency selectivity. Typically, they manipulate signals in momentum space for optical edge detection imaging at the target wavelength regime {\cite{Guo2018,kwon2018,Kwon2020, Komar2021,zhou2024}}. Recent studies have shown that the introduction of quasi-bound states in the continuum (Q-BIC) into nonlocal metasurfaces is accelerating the field's evolution from fundamental wavefront shaping toward multifunctional integration, intelligent responsiveness, and practical applications {\cite{Zhou2020a,Pan2021,liu2024,zhou2025,Zhang2025,Zong2025}}. Initial breakthroughs demonstrated that breaking structural symmetry can activate high-Q Q-BIC modes, enabling efficient beam manipulation and metalens imaging {\cite{overvig2020}}. To overcome the spectral limitations of single-layer metasurfaces, researchers have explored multi-wavelength modulation strategies using multilayer structure and multi-parameter disturbance designs{\cite{malek2020,malek2022}}. This ability to customize wavefronts with multiple degrees of freedom provides a physical foundation for joint spectral–spatial encoding. However, the efficiency of polarization conversion for the geometric phase modulation of Q-BIC remains limited. Then a single-layer nonlocal Huygens’ meta-lens achieves a transmission polarization conversion efficiency exceeding the theoretical limit of 25\%{\cite{yao2024}}. Further advances utilize converted and unconverted light components to enable multifunctional imaging including bright-field imaging and edge extraction{\cite{yao2025}}. However, current nonlocal metasurface designs still face two major challenges: (1) The inability to simultaneously extract edge features and depth-of-field information, limiting applicability in 3D scene perception; and (2) The lack of reconfigurability in existing devices, hindering their suitability for dynamic optical computing.

In this work, we propose a wavelength-tunable nonlocal metasurface that simultaneously extracts image depth information and edge features. Enabled by Q-BIC and magnetic dipole resonance (MDR) coupling, this approach overcomes the efficiency limitations inherent in conventional circular polarization conversion, achieving over 40\% efficiency. By leveraging the dynamic modulation of the phase-change material refractive index Sb$_{2}$S$_{3}$, the system achieves continuous resonance wavelength tuning across a 100 nm range. Through spin-multiplexing, parallel dual-function processing is realized: the converted LCP component undergoes the Pancharatnam-Berry (PB) phase-based wavefront shaping to establish a quantitative mapping relationship between focal rotation angle and depth information, extracting depth data; the unconverted RCP component utilizes the metasurface's nonlocal characteristics for momentum-space filtering to extract edge features. This integrated optical design provides an innovative solution for applications requiring real-time depth perception and feature recognition, such as autonomous driving systems and biomedical imaging.

\section{Results and discussion}

\subsection{Metasurface design principle }

\begin{figure*}
	\centering % 使图片居中
	\includegraphics[width=\textwidth]{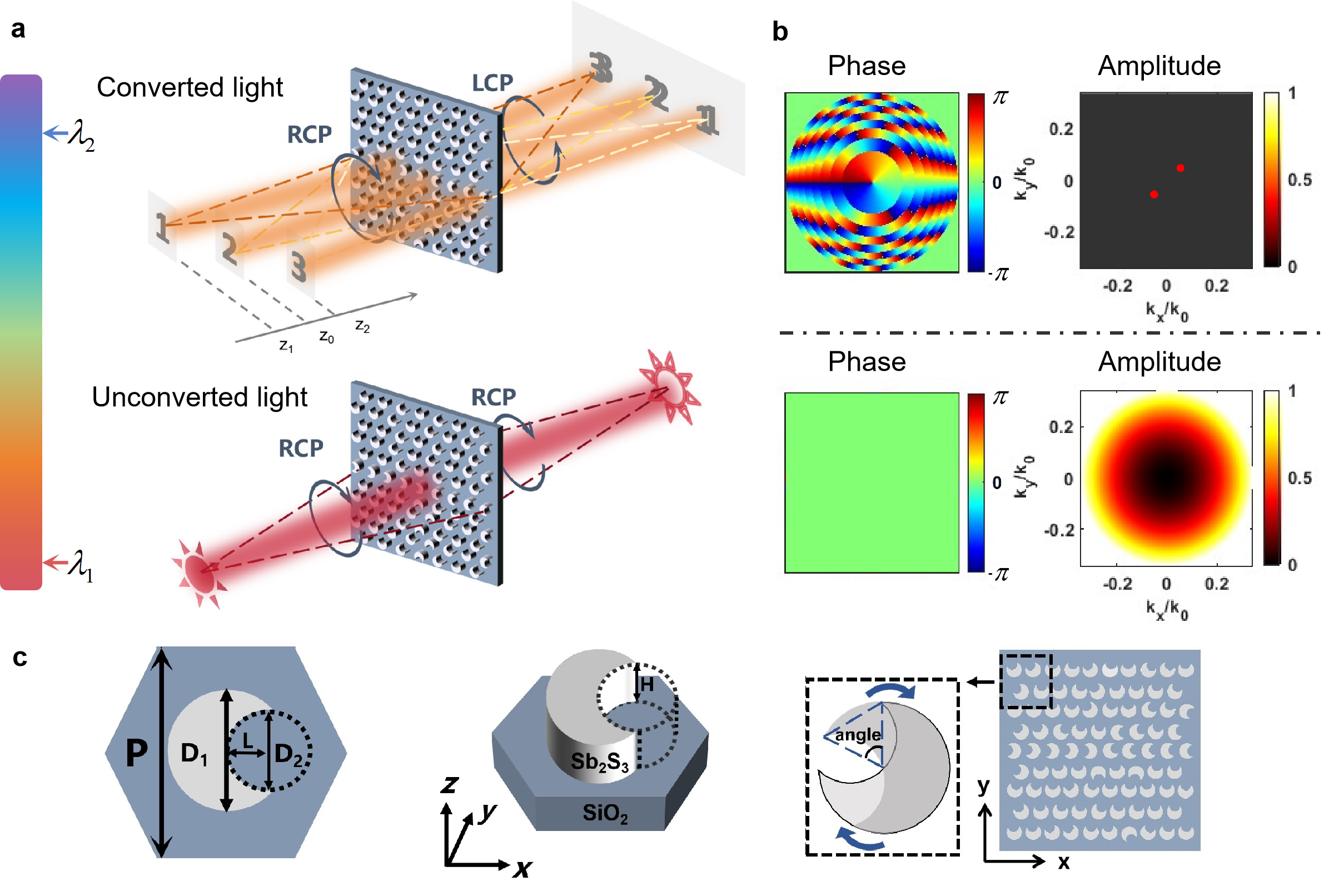} % 使用 \textwidth 确保图片宽度填满整个页面
	\caption{\label{Fig. 1}Schematic illustration of the wavelength-tunable nonlocal Huygens’ metasurface for spin-multiplexed edge-enhanced imaging and depth sensing. (a) Under right-circularly polarized (RCP) illumination, the converted left-circularly polarized (LCP) light, modulated by the metasurface's geometric phase profile, outputs the computationally extracted image depth information. The unconverted RCP component undergoes momentum-space filtering by the metasurface, generating an edge-enhanced image. The color bar on the left indicates the resonant wavelength shift associated with the refractive index change of Sb$_{2}$S$_{3}$. (b) Phase and amplitude profiles corresponding to RCP and LCP for spin-multiplexed imaging. (c) Metasurface unit cell comprising crescent-shaped Sb$_{2}$S$_{3}$ nanopillars deposited on a glass substrate in a hexagonal array. The rotation angle of the nanopillars imparts a geometric phase gradient.} % 图片标题
\end{figure*}

Fig. \ref{Fig. 1}(a) illustrates a wavelength-tunable nonlocal Huygens’ metasurface that enables spin-multiplexed edge-enhanced imaging and depth sensing. Upon transmission of the converted LCP component, distinct axial displacements represented by the digital inputs "1", "2", and "3" generate two separated focal points on the output plane in the converted LCP component. Generating these distinct focal spots corresponding to different axial positions requires a vortex phase profile with multiple topological charges. Depth information extraction is achieved by establishing a mapping relationship between the azimuthal angle of the separated images and the axial displacement of the incident image. This requires sufficient transmission intensity and high imaging fidelity for the converted LCP component. Simultaneously, the transmitted unconverted RCP component, which carries the "sun" image information, yields an edge-enhanced image on the output plane. This edge enhancement is achieved by leveraging the nonlocal properties of the metasurface: it exhibits near-zero transmission at low incident angles, effectively filtering out low-spatial-frequency components and thus enabling edge detection. Fig. \ref{Fig. 1}(b) presents the amplitude and phase responses enabling multifunctional operation. To overcome the intrinsically limited wavelength selectivity of nonlocal metasurfaces, the phase-change material Sb$_{2}$S$_{3}$ is incorporated, enabling versatile image processing at multiple selectable wavelengths, with the incident wavelength continuously tunable across a designated spectral range. The transmittance ratios \textit{T$_{RL}$} (for LCP) and \textit{T$_{RR}$} (for RCP) are defined as the ratio of the output optical power of LCP (RCP) to the incident optical power of RCP. Conventional metasurface designs cannot simultaneously achieve high \textit{T$_{RL}$} and low \textit{T$_{RR}$}, as they are often constrained by the theoretical maximum polarization conversion efficiency limit of 25\%. Consequently, a specialized unit cell structure must be engineered to overcome this limitation.

Fig. \ref{Fig. 1}(c) presents the unit cell structure constituting the metasurface. Crescent-shaped Sb$_{2}$S$_{3}$ nanopillars with height $H = 327$ nm sit on a glass substrate arranged in a hexagonal lattice with period $P = 1000$ nm. Each nanopillar is formed by etching a smaller cylindrical hole of diameter $\textit{D$_{2}$} = 500$ nm, offset by $L = 220$ nm, from a larger cylindrical pillar with diameter $\textit{D$_{1}$} = 620$ nm. As schematized in Fig. \ref{Fig. 1}(c), the modulation of the PB phase is encoded by rotating the crescent orientation angle. For the output unconverted RCP light, the phase response of the nonlocal metasurface remains invariant with respect to the unit cell orientation angle, whereas the amplitude response exhibits strong nonlocal dispersion characteristics, which is essential for the edge detection functionality. For the converted LCP output, the metasurface phase corresponds to the multiplexed vortex phase profiles with multiple topological charges, and the amplitude map reveals the rotating focal spots generated by the phase distribution, enabling DoF detection.

\begin{figure}
	\centering % 使图片居中
	\includegraphics[width=\linewidth]{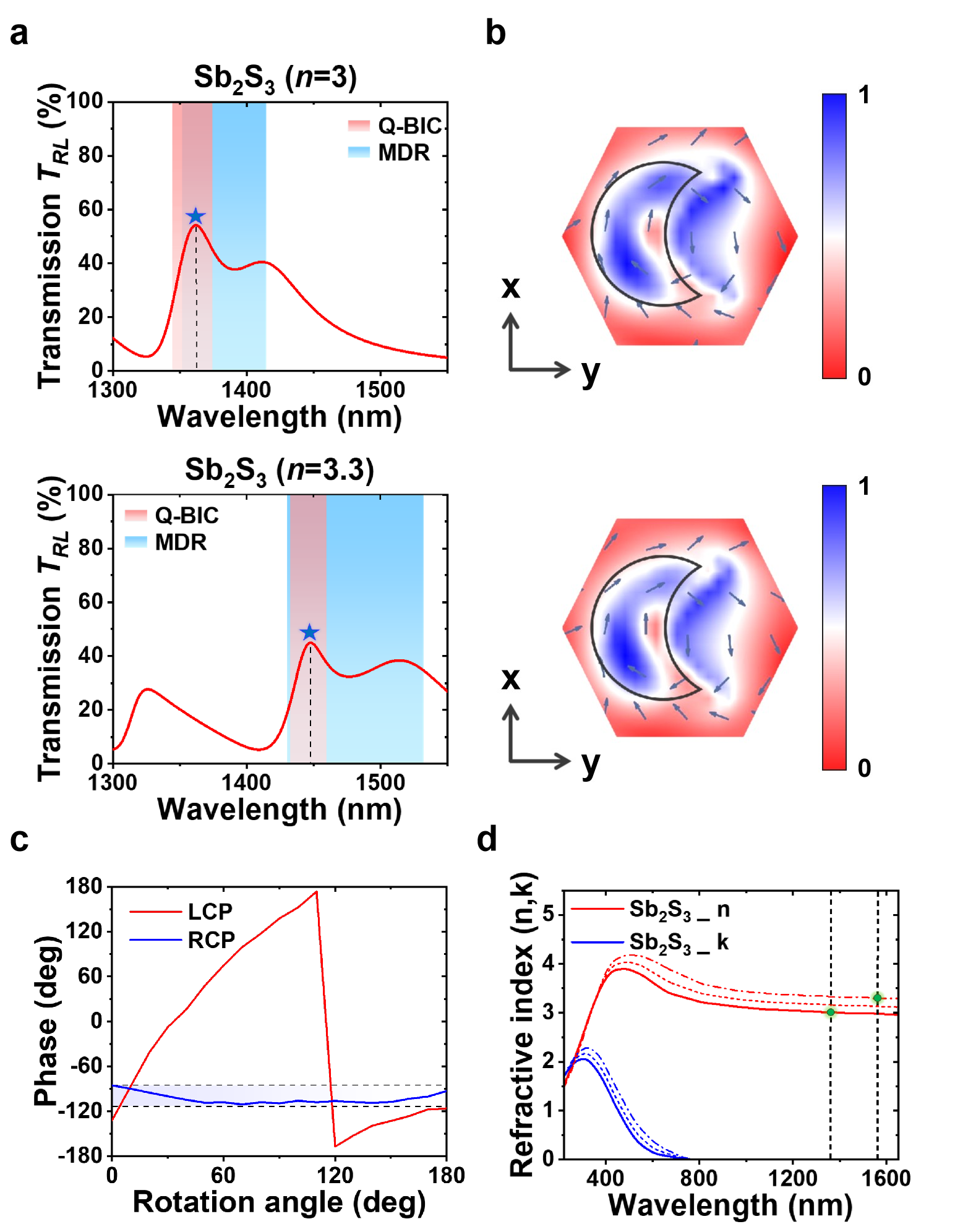} % 使用 \textwidth 确保图片宽度填满整个页面
	\caption{\label{Fig. 2}Optical responses of output converted LCP light. (a) Transmission spectra of the converted LCP component for Sb$_{2}$S$_{3}$ metasurfaces with refractive indices of 3.0 and 3.3. The red and blue shaded regions denote the bandwidths of the Q-BIC and MDR, respectively. (b) Electric field and current vector profiles within the unit cell at the resonance wavelengths for metasurfaces with the corresponding refractive indices. (c) Phase profiles of the metasurface for output RCP and LCP light. (d) Spectral variation of the refractive index (n, k) of Sb$_{2}$S$_{3}$.} % 图片标题
\end{figure}

Fig. \ref{Fig. 2}(a) illustrates the transmission spectra of \textit{T$_{RL}$} of the designed nonlocal metasurfaces across 1300 nm to 1550 nm range under different refractive indices of Sb$_{2}$S$_{3}$ nanopillars, quantifying its high conversion efficiency. Due to structural asymmetry, the symmetry-protected Q-BIC modes are excited under different refractive indices of Sb$_{2}$S$_{3}$ nanopillars. For Sb$_{2}$S$_{3}$ with $n = 3.0$, a narrow Q-BIC resonance (red region) at 1368 nm spectrally overlaps with a transmission peak (blue region) of magnetic dipole resonance (MDR) at 1380 nm. Their interference generates a Fano lineshape and satisfies the generalized Kerker condition for forward scattering, achieving the transmission amplitude at the \textit{T$_{RL}$} peak reach approximately 55\%. The similar case applies to the metasurface with Sb$_{2}$S$_{3}$ of $n = 3.3$, where the excitation of the Q-BIC mode at 1455 nm and the MDR at 1475 nm couples to reach a peak \textit{T$_{RL}$} of 45\%. Analysis of the in-plane electric field in the unit cell, shown in Fig. \ref{Fig. 2}(b), revealed by white vector arrows, confirms magnetic dipole resonance at these working wavelengths through closed-loop circulating currents. 
Fig. \ref{Fig. 2}(c) shows that the PB phase rotation angle spans from 0° to 180°, covering a full geometric phase range from -$\pi$ to $\pi$ for the converted LCP component with high \textit{T$_{RL}$}. Meanwhile, the phase of the unconverted RCP component remains nearly invariant within this rotation angle range. Fig. \ref{Fig. 2}(d) shows the variation in the refractive index of  Sb$_{2}$S$_{3}$ ($n$, $k$) in the wavelength regimes of interest, where two green dots represent the two distinct refractive indices of Sb$_{2}$S$_{3}$, corresponding to working wavelengths 1368 nm(Dip1) and 1455 nm(Dip2). This confirms that, as the refractive index varies, the operating wavelength of the metasurface also continuously shifts within this range. The combination of high transmission \textit{T$_{RL}$} and stable geometric phase of the output LCP component facilitate high-performance multi-wavelength wavefront modulation for DoF detection. In particular, integrating Sb$_{2}$S$_{3}$ as a nonvolatile phase-change material overcomes the static limitations of conventional metasurfaces. Unlike electro-optic approaches requiring sustained power {\cite{buchnev2015,xiao2020,li2022,momeni2022switchable}}, Sb$_{2}$S$_{3}$ enables nanosecond switching with large index contrast ($n > 0.8$) and non-volatile multi-states {\cite{zhou2020,zhang2021}}. This facilitates dynamic reconfigurability for various optical functions, including structural color generation {\cite{pan}}, holographic imaging {\cite{moitra2023programmable,liu2024a}}, and optical edge detection {\cite{cotrufo2024,guo2025,liu2025,Yang2025}}. Though controlling the crystallization fraction during phase transition, active tunable resonance wavelengths of the proposed nonlocal metasurfaces can be achieved, overcoming their inherent narrowband limitations.

\begin{figure}
	\includegraphics[width=\linewidth]{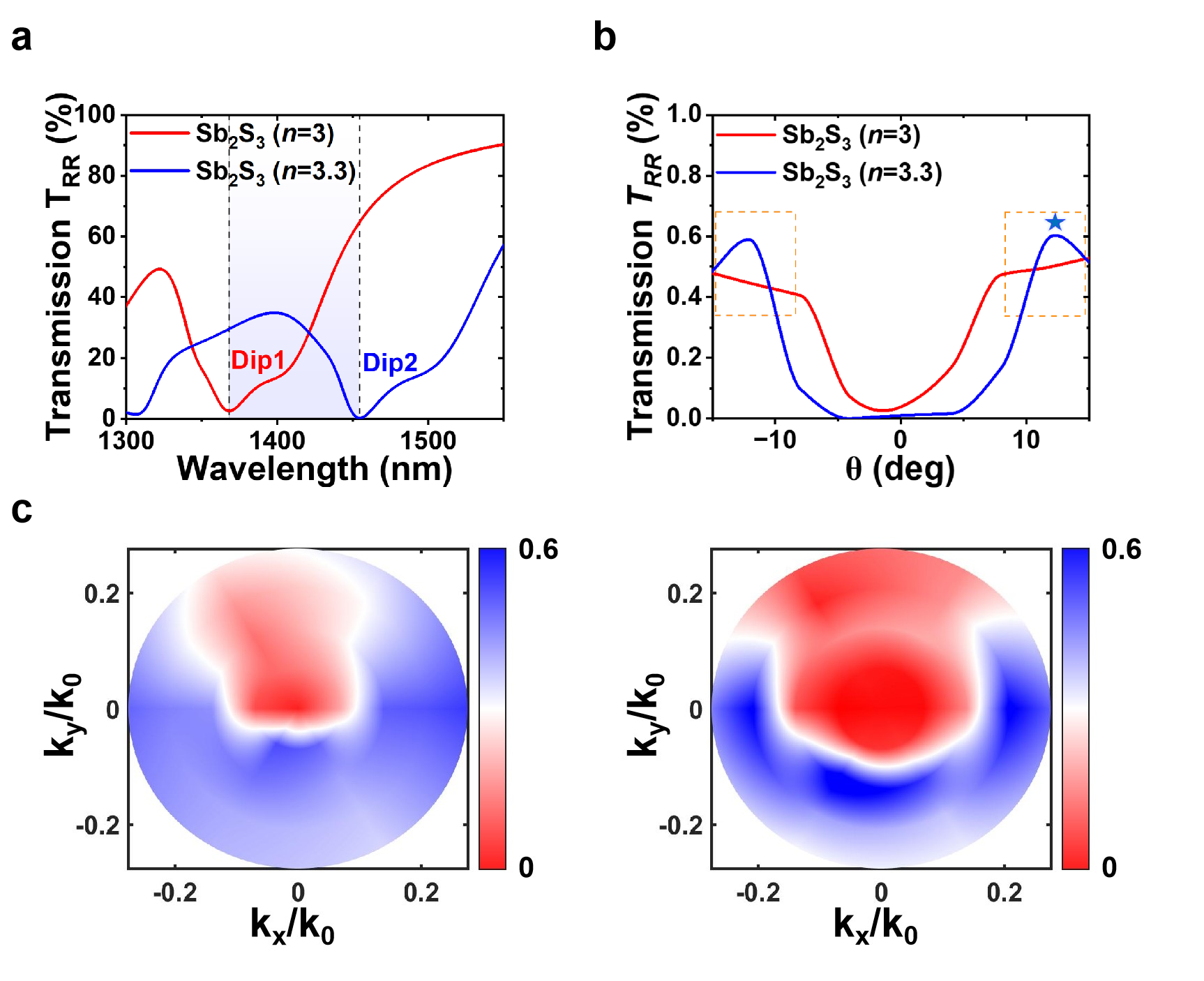}% Here is how to import EPS art
	\caption{\label{Fig. 3}Optical responses of output unconverted RCP light. (a) Transmission spectra of the Sb$_{2}$S$_{3}$ metasurface with refractive indices varying from 3 to 3.3. (b) Transmittance variation with incident angle for Sb$_{2}$S$_{3}$ metasurfaces ($n = 3$ and $n = 3.3$) at their respective resonant wavelengths. (c) Two-dimensional transmission dispersion maps at the operating wavelengths for Sb$_{2}$S$_{3}$ metasurfaces with refractive indices of 3 and 3.3, respectively.} % 图片标题
\end{figure}

On the other hand, the transmission of the unconverted output light, i.e. the RCP component, is provided in Fig. \ref{Fig. 3}(a). The resonance wavelengths for metasurfaces with refractive indices of 3 and 3.3 are located at 1368 nm and 1455 nm, respectively, exhibiting minimal transmission at normal incidence. As the refractive index  of Sb$_{2}$S$_{3}$ varies, the resonance wavelength shifts continuously between 1368 nm and 1455 nm. This designed Sb$_{2}$S$_{3}$ metasurface with wavelength selectivity over a nearly 100 nm bandwidth exceeds the performance of conventional nonlocal metasurfaces operating effectively only within a limited spatial range. Fig. \ref{Fig. 3}(b) illustrates the quadratic dependence of the transmission coefficient on the incident angle varying from 0° to 16°, for metasurfaces at different refractive indices. Under oblique incidence, a redshift of the resonance wavelength occurs, leading to suppressed transmission at normal incidence and gradually increasing transmission with increasing angle. The maximum transmission modulation reaches 52\% and 60\% for the two refractive indices, respectively. This pronounced angular dispersion arises from interactions between adjacent meta-atoms, which is a hallmark of nonlocal metasurface. Thus, the nonlocal dispersion effect filters out the central low spatial frequencies and transmits the high spatial frequencies, enabling image edge detection. To more intuitively illustrate this high-pass filtering effect, transmission spectra are simulated across incident angles ranging from 0° to 16° and azimuthal angles from 0° to 360°, as shown in Fig. \ref{Fig. 3}(c). The optical transfer functions clearly show near-zero transmission for low-frequency components and enhanced transmission for high-frequency components. These characteristics, including wavelength-selective resonance, strong angular dispersion, and effective high-pass filtering, enable the unconverted RCP component of the metasurface to perform tunable two-dimensional image edge detection.

\subsection{Image processing of the proposed nonlocal metasurface}

\begin{figure}
    \centering % 使图片居中
    \includegraphics[width=\linewidth]{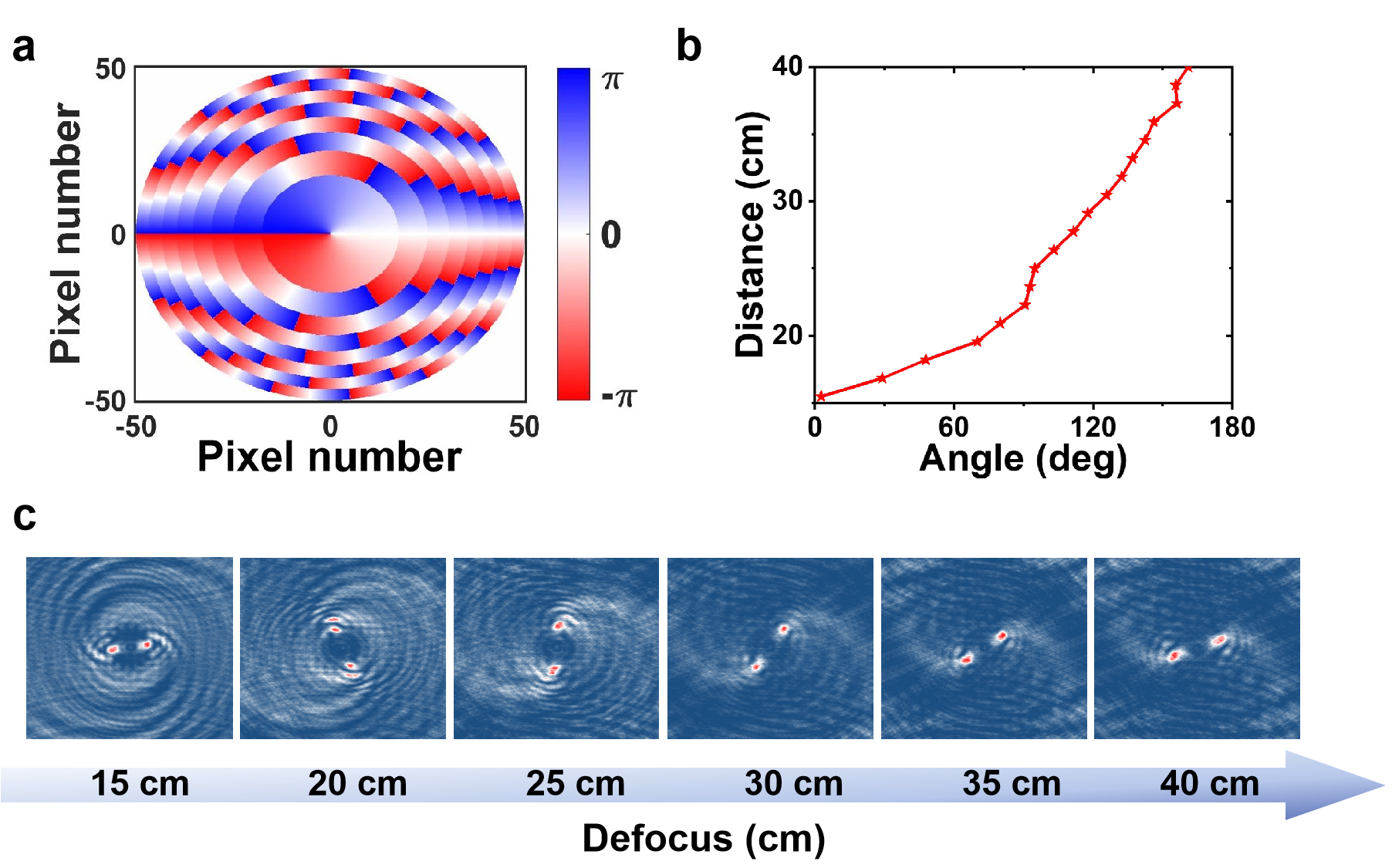} % 使用 \textwidth 确保图片宽度填满整个页面
    \caption{\label{Fig. 4}Design of the metasurfaces for depth sensing. (a) Phase profile of multiplexed vortex phase profiles with multiple topological charges. (b) Mapping curve between the azimuthal angle of the twin images and the axial defocus displacement of the incident image. (c) Schematic illustration of the two focal spots’ variation at diffraction distances of 15 cm, 20 cm, 25 cm, 30 cm, 35 cm, and 40 cm.} % 图片标题
\end{figure}

We employ a spiral phase (SP)-based method to generate a double-helix point spread function (DH-PSF) phase profile for depth sensing. By varying the number of Fresnel zones, the rotation rate of the DH-PSF can be easily controlled. These favorable characteristics allow the DH-PSF’s to adjust to the axial range and rotational speed by adjusting the relevant parameters. The phase distribution satisfies Eq. (\ref{eq:one}),
\begin{equation}
\phi = 
\begin{cases} 
(2n - 1)\varphi, & R\sqrt{\dfrac{n - 1}{N}} < \rho < R\sqrt{\dfrac{n}{N}}, \\
& \quad n = 1, 2, \ldots, N \\ % 调整间距
0, & \rho > R 
\end{cases}
\label{eq:one}.
\end{equation}
where $\phi$ denotes the phase composed of multiple Fresnel zones. The maximum radius is $R = 50$ and the number of Fresnel zones is $N$. Each Fresnel zone corresponds to a vortex beam with a topological charge that increases by 2 in arithmetic progression from the innermost to the outermost zone. Considering the detectable DoF range and precision requirements, $N = 8$ is selected for the phase distribution shown in Fig. \ref{Fig. 4}(a). The stable geometric phase distribution shown in Fig. \ref{Fig. 2}(c) ensures similar responses between the unit cells and the meta-lens. The propagation of light and the response to the metasurface are modeled according to Eqs. (\ref{eq:two}) and (\ref{eq:three}),
\begin{align}
A(f_x, f_y; 0) &= \iint f(x, y) \cdot \exp(i\varphi) \notag \\
&\quad \exp\left(-i2\pi(xf_x + yf_y)\right)  dx  dy
\label{eq:two}.
\end{align}

\begin{align}
U(x, y, z) &= \iint A(f_x, f_y; 0) \cdot \exp\left(i2\pi(xf_x + yf_y)\right) \notag \\
&\quad \exp\left(i2\pi z \sqrt{\dfrac{1}{\lambda^2} - f_x^2 - f_y^2}\right)  df_x  df_y
\label{eq:three}.
\end{align}
Here, $A(f_x,f_y,0)$ represents the complex amplitude in the Fourier plane after modulation by the PB phase profile. Angular spectrum propagation is then applied to compute the complex amplitude $U(x,y,z)$ at the image plane in Eq. (\ref{eq:three}). 
Consequently, by encoding the phase profile according to Eq. (\ref{eq:one}), a mapping between the object depth and the focal spot rotation angle is established, as depicted in Fig. \ref{Fig. 4}(b). Within a rotation range of 0° to 180° , the detectable depth range spans from 15 cm to 40 cm. Angular spectrum propagation is employed to simulate the diffraction process, yielding the detailed evolution of the focal spot rotation shown in Fig. \ref{Fig. 4}(c). For depths ranging from 15 cm to 40 cm in 5 cm increments, the corresponding rotation angles of the defocused focal spots are measured and validated. These results confirm that the measured depth-angle relationship matches the calibration curve presented in Fig. \ref{Fig. 4}(b).

\begin{figure}
    \includegraphics[width=\linewidth]{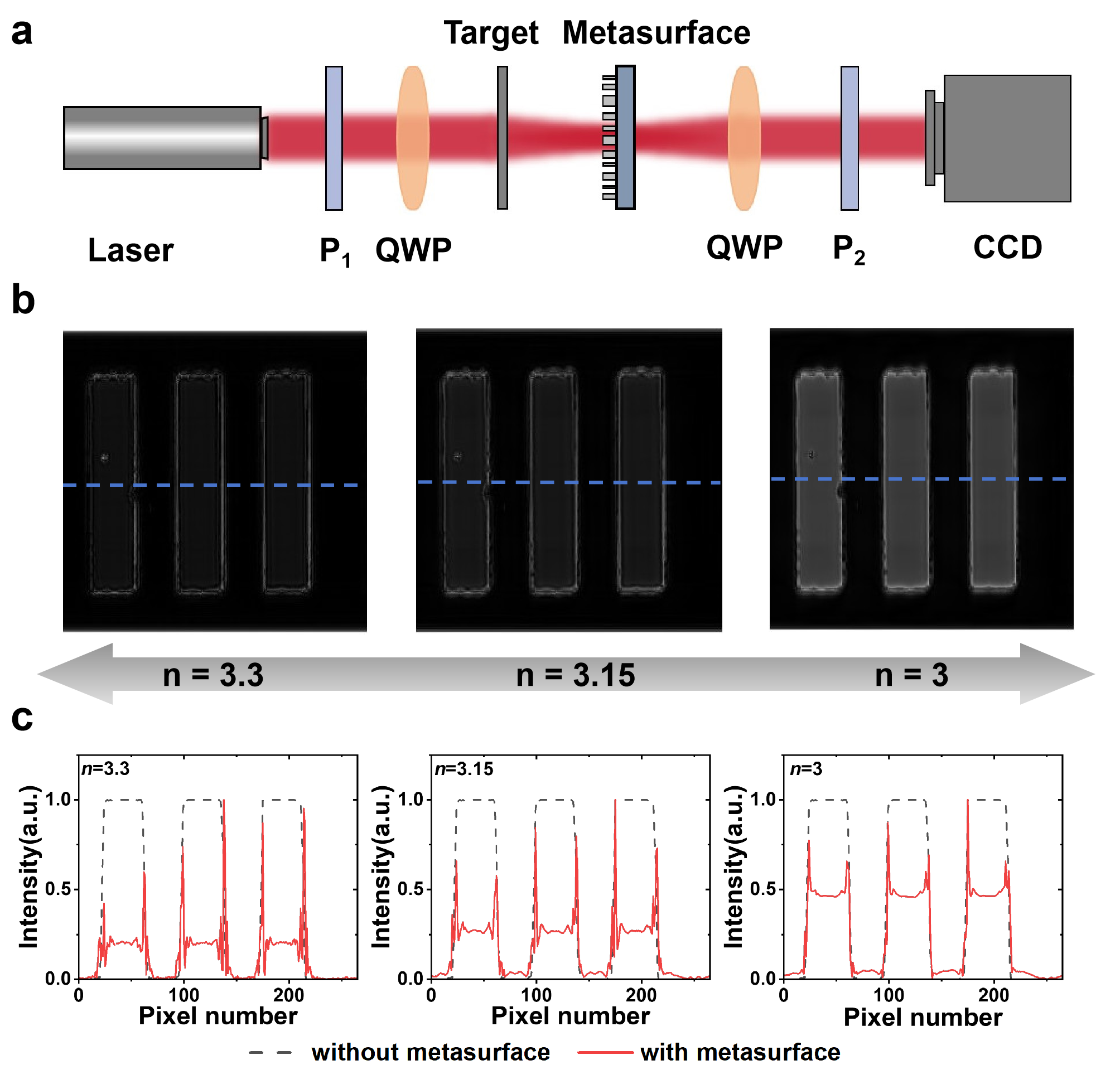} % 使用 \textwidth 确保图片宽度填满整个页面
    \caption{\label{Fig. 5}Performance of the metasurface for edge detection using a resolution target. (a) Schematic diagram of the experimental setup. Laser: light source; \textit{P$_1$} and \textit{P$_2$}: polarizer and analyzer; QWP: quarter-wave plate; Target: resolution mask; CCD: imaging sensor. Incident light passes through the designed metasurface, and by rotating the analyzer, different optical functionalities are captured by the CCD. (b) Edge-enhanced images obtained from unconverted light after transmission through metasurfaces with Sb$_{2}$S$_{3}$ refractive indices of 3.3, 3.15, and 3, respectively. (c) Cross-sectional intensity profiles along the blue dashed lines in (b), showing edge contrast performance under three refractive indices.} % 图片标题
\end{figure}

Fig. \ref{Fig. 5}(a) shows the system configuration in which incident light is modulated by the metasurface and the corresponding function is retrieved on the CCD. To quantitatively evaluate the edge detection performance of the metasurface, we employed the 1951 USAF Resolution Test Chart. As shown in Fig. \ref{Fig. 5}(b), edge detection images are obtained with refractive indices $n=3.3$, 3.15 and 3. The left image ($n =3.3$) shows clear and sharp edges, indicating effective edge enhancement. The right image ($n =3$) also reveals enhanced edges, although the contrast is slightly reduced. The middle image corresponds to the intermediate refractive index $n =3.15$. Fig. \ref{Fig. 5}(c) presents the horizontal cross-sectional intensity profiles along the blue dashed lines in Fig. \ref{Fig. 5}(b), where the x-axis represents pixel positions across the resolution target. The intensity distribution along the horizontal axis reveals distinct intensity peaks at the edges of the target features, indicating effective enhancement by the Laplacian operator. The metasurface clearly suppresses low-frequency components and improves image contrast. As the refractive index decreases from 3.3 to 3, reduced suppression of low-frequency components, leads to a slight reduction in contrast. Nonetheless, the edge detection functionality is well preserved throughout this range.

\begin{figure}
	\centering % 使图片居中
	\includegraphics[width=\linewidth]{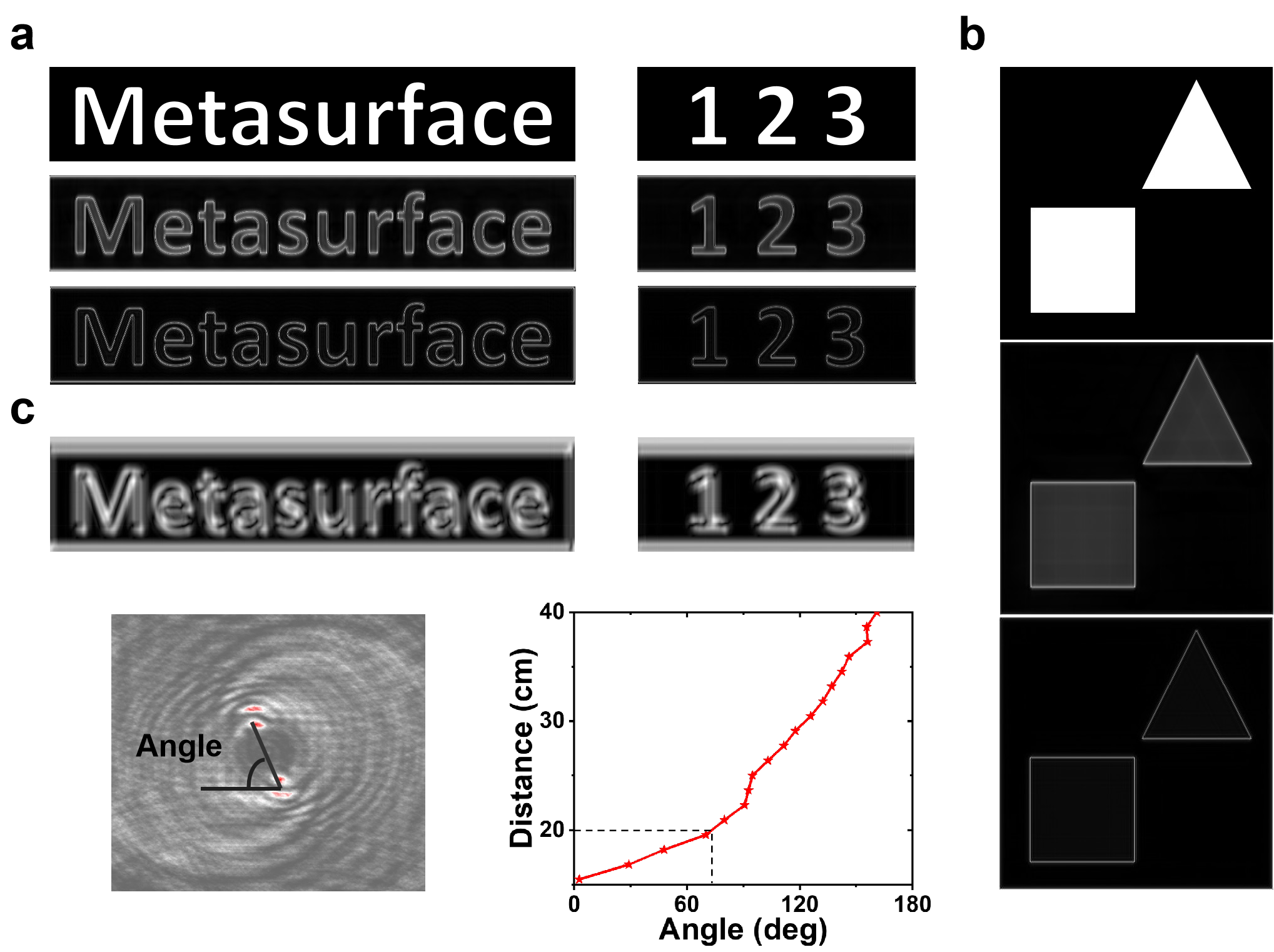} % 使用 \textwidth 确保图片宽度填满整个页面
	\caption{\label{Fig. 6}Image processing performance of the metasurface. (a) Input images "Metasurface" and "123", and their corresponding edge detection results after transmission through metasurfaces with refractive indices $n =3.3$ and $n =3$, respectively. (b) Edge detection results for square and triangular input images after passing through the two metasurfaces. (c) A pair of defocused images captured under a depth of field of 20 cm, corresponding to overlapping features with angular separation.} % 图片标题
\end{figure}

To validate the imaging capabilities of the metasurface, input images featuring the words "Metasurface", "123", along with square and triangular patterns, are utilized in simulations. As illustrated in Figs. \ref{Fig. 6}(a) and \ref{Fig. 6}(b), the original input images and the corresponding edge-enhanced outputs modulated by metasurfaces with refractive indices of $n=3.3$ and $n=3$ are shown. The edges of the objects are clearly delineated, demonstrating effective edge enhancement functionality. Fig. \ref{Fig. 6}(c) presents the DoF output, where two defocused images partially overlap. The two subfigures below Fig. \ref{Fig. 6}(c) depict the azimuth angle of these two defocused images. By calculating the azimuth angle separation between the overlapping features and referencing the pre-calibrated angle-depth mapping curve, the corresponding object depth is determined to be 20 cm. These results demonstrate that the metasurface achieves dual functionality: high-fidelity optical edge detection and depth-resolved imaging. Crucially, as the phase-change material Sb$_{2}$S$_{3}$ shifts its refractive index from 3.0 to 3.3, the unconverted RCP component transitions from edge-enhanced images to outputs with further accentuated edge contrast. Throughout this progress, the performance of the converted LCP component in computational depth extraction remains consistent. This design integrates DoF and optical edge detection functionalities within a single metasurface. By physically encoding both spatial features and depth information at the optical hardware level, this technology establishes a foundation for applications requiring integrated scene analysis—including medical pathology localization imaging and minimally invasive endoscopic diagnostics—while reducing computational overhead.

\section{Conclusions}
In conclusion, we have demonstrated that a nonlocal Huygens’ metasurface can simultaneously extract depth information and edge features. Specifically, the converted LCP component utilizes a specially designed vortex phase profile based on the PB phase. This establishes a mapping relationship between the azimuthal angle of two separated images and the axial displacement of the input image, thereby enabling the acquisition of the corresponding depth information. Meanwhile, the unconverted RCP component leverages the nonlocal characteristics of the metasurface to perform momentum-space filtering, extracting the edge features of the image. By exploiting the coupling between Q-BIC and MDR, the metasurface satisfies the generalized Kerker condition, overcoming the efficiency limitations inherent in conventional circular polarization conversion. Furthermore, the integration of the phase-change material Sb$_{2}$S$_{3}$, which possesses a tunable refractive index, extends the spectral selectivity range of the nonlocal resonant metasurface, significantly enhancing its practical applicability. This multifunctional imaging system exhibits significant potential for applications in fields such as autonomous vehicle navigation and biomedical lesion imaging.

\begin{acknowledgments}
This work was supported by the National Natural Science  Foundation of China (Grants No. 12304420, No. 12264028, and No. 12364045), the Natural Science Foundation of Jiangxi Province (Grants No. 20232BAB201040 and No. 20232BAB211025), and the Young Elite Scientists Sponsorship Program by JXAST (Grants No. 2023QT11 and 2025QT04).
\end{acknowledgments}

% \bmsection{Supplemental document}
% See Supplement 1 for supporting content.

\nocite{*}

%\bibliography{Ref}% Produces the bibliography via BibTeX.


\end{document}